\documentclass[aps,twocolumn,amsfonts,amssymb,amsmath,showpacs,letterpaper,superscriptaddress,nofootinbib,altaffilletter,fontenc, floatfix, 10pt]{revtex4-2}

\usepackage{float}
\usepackage{epsfig}
\usepackage{url}
\usepackage{multirow}
\usepackage{epstopdf}
\usepackage{graphicx}
\usepackage{orcidlink}
\usepackage{ragged2e}
\usepackage{caption}
\usepackage{ulem}
\usepackage{subfigure}
\usepackage{latexsym}
\usepackage{comment}
\usepackage{array}
\usepackage[dvipsnames]{xcolor}

\makeatletter
\def\@float@table{%
  \def\fs@plain{%
    \def\@fs@cfont{\bfseries}%
    \let\@fs@capt\relax
    \def\@fs@pre{}%
    \def\@fs@post{}%
    \def\@fs@mid{\kern10pt}%
    \let\@fs@iftopcapt\iftrue
  }%
  \float@plain
}

\def\fnum@table{Table~\thetable}
\makeatother
\usepackage{caption}
\usepackage{booktabs}
\makeatletter
\long\def\@makecaption#1#2{%
  \vskip\abovecaptionskip
  \sbox\@tempboxa{\small #1. #2}%
  \ifdim \wd\@tempboxa > \hsize
    \small #1. #2\par  
  \else
    \global\@minipagefalse
    \hb@xt@\hsize{\hfil\box\@tempboxa\hfil}%
  \fi
  \vskip\belowcaptionskip
}
\makeatother

\begin{document}

\title{ Bayesian parameter estimation for the Core-bounce phase of Rapidly Rotating Core-Collapse Supernovae in real interferometric data
}


\author{Emmanuel Avila} 
\email[E-mail: ]{emmanuel.avila@tec.mx}
\affiliation{Tecnologico de Monterrey, Escuela de Ingeniería y Ciencias, Monterrey, N.L., 64849, México}

\author{Michele Zanolin} 
\email[E-mail: ]{zanolinm@erau.edu}
\affiliation{Embry-Riddle Aeronautical University, Prescott, AZ 86301, USA}

\author{Javier~M.~Antelis}
\email[E-mail: ]{mauricio.antelis@tec.mx}
\affiliation{Tecnologico de Monterrey, Escuela de Ingeniería y Ciencias, Monterrey, N.L., 64849, México}

\author{Claudia Moreno} 
\email[E-mail: ]{claudia.moreno@academico.udg.mx}
\affiliation{Departamento de F\'isica, CUCEI, Universidad de Guadalajara
C.P. 44430, Guadalajara, Jal., M\'exico}


\date{\today}

%
\bigskip
\begin{abstract}
In this work, we present a novel methodology for estimating the ratio of kinetic energy to gravitational potential energy of a core collapse supernova progenitor and assessing the Equation Of State (EOS). We use the reconstruction of a gravitational wave produced by the core bounce phase of collapse supernovae in rapidly rotating progenitors in real interferometric data. For this purpose, we extend a previous phenomenological analytical model for the core bounce phase by introducing an additional parameter associated with its timescale. We assess the agreement of our phenomenological template with numerical waveform databases through fitting factor and Bayesian model comparison. We also quantify the consistency of the databases between each other. The improved phenomenological analytical model raised the median fitting factors from 88.88\% to 90.83\%. Parameter estimation (PE) is performed using a Markov Chain Monte Carlo implementation in real O3aL1 \cite{abbott2022ligoO3} interferometric noise. We find that the rotational parameter $\beta$ estimation for 452 Abylkairov signals \cite{abylkairov2025assessing} has a median absolute relative error of 11.93\% with a 95th percentile of 38.41\%, an overall uncertainty of $\sigma_\beta = 1.083 \times 10^{-3}$ at 10 kpc. This was an improvement compared with estimated values  using matched filtering and Maximum Likelihood Estimation (MLE) \cite{laura}, in which at 10 kpc, a value of $\sigma_\beta = 1.46\times 10^{-2}$ was reported. We further investigate the impact of the bayesian prior selection on the posteriors, injecting signals in gaussian colored noise based on LIGO PSD and real interferometric noise. 
Using real noise would introduce non-gaussian and non-stationary features that worsen the estimation accuracy, for example, the estimation for $\beta$, yielded a maximum bias of 11.9\% at 10 kpc for a uniform in $\beta^2$ prior and a minimum bias of 0.6\% with a triangular prior.

\end{abstract}

\maketitle

\section{Introduction}
\label{Sec:Introduction}

Core-Collapse Supernovae (CCSNe) are one of the primary targets of Gravitational Wave (GW) astronomy.  The detectability of GWs emitted by CCSNe has been studied by the LIGO Virgo and Kagra collaborations: see for example a discussion of detection scenarios in \cite{gossan2016observing}, multi-measurement analysis strategies in \cite{halim2021multimessenger}, a deep learning perspective in \cite{iess2020core}, and a general overview of the literature in \cite{mezzacappa2024gravitational}.

Several groups developed numerical simulations for slowly rotating and rapidly rotating progenitors (see, for example, \cite{couch2014high,andresen2017gravitational,summa2018rotation} \cite{dimmelmeier2008gravitational,fuller2015supernova,powell2020three} \cite{schnauck2026gravitational,shibagaki2024three} and for a review \cite{mezzacappa2020toward}).
The GW signal starts during the core bounce phase, followed by the accretion stage, and in many aspects, the signal is stochastic. Nevertheless, there are deterministic features for non-rotating and rotating models in both the time and frequency domains like (a) the core bounce in rapidly rotating progenitors (as was pointed out in \cite{richers2017equation} and \cite{janka2012core,mezzacappa2020physical,murphy2025core,mezzacappa2024gravitational}), (b) the evolution of the resonant frequency in the fundamental f and g modes of the Proto Neutron Star (PNS), (c) the standing accretion shock instability (SASI) \cite{blondin2003stability}  and (d) memory  \cite{morozova2018gravitational,torres2019universal,mezzacappa2024gravitational}. 
This work focuses on parameter estimation from the core bounce component of Rapidly-Rotating CCSN.

For Rapidly-Rotating CCSN the core bounce GW generation occurs during the first 10 ms, followed by ring-down oscillations of the forming PNS. The axisymmetric conditions of the initial perturbations of the PNS produces an elliptically polarized gravitational wave, that enables the use of 2D simulations for extracting the $h_{+}$ polarization of the GW signal during bounce \cite{abdikamalov2014measuring}. The amplitude of the signal in this stage is determined mainly by the rotational ratio $\beta = T/|W|$ between the rotational kinetic energy $T$ and the gravitational binding energy $|W|$ \cite{shibagaki2020new, rover2009bayesian}. The equation of state is expected to have a role as well \cite{richers2017equation}.

Previous parameter estimation investigations applied Bayesian inference to identify physical quantities of the progenitor star and the PNS during the bounce. For instance, in Röver et al. (2009) \cite{rover2009bayesian} used Principal Component Analysis in simulated gaussian noise to estimate the rotational parameter at the time of the bounce, the post-bounce oscillations, and the maximum density of the core also at bounce ($\rho_b$). Those parameters provide information about the dynamics of the collapse and whether the bounce is pressure-dominated (occurs via EOS stiffening) or is dominated by the centrifugal support provided to the core through rotation. 

During the collapse of the inner core, the EOS can provide support against gravity, along with the rotation rates, which can delay or halt the bounce due to centrifugal support. According to the Teukolsky \& Shapiro \cite{shapiro2024black} about the physics of collapsing compact objects, for a non-static star we can estimate the free-fall time when there is no pressure to support it against gravitational collapse, and moreover, if it oscillates or expands. This timescale is inversely proportional to $\sqrt{G\rho}$. Following the results yielded by the principal component analysis from Röver et al. (2009) and the fact that the bounce is a contraction followed by a sudden expansion, the bounce duration is $T_b = 1/f_b$ and it can be measured through the frequency $f_b = \sqrt{G\rho_b}$. That is given by the gravitational constant G and $\rho_b$, the density of the inner core at bounce.

Richers et. al. generated a dataset of around 1800 2D axisymmetric  general-relativistic hydrodynamic simulations covering a parameter space of 98 rotation profiles and 18 Equations of State (EOS) and showed the impact of them in the core bounce and post-bounce dynamics. Also, Mitra, et al. (2023) \cite{mitra2023exploring} released a catalog of 402 2D waveforms generated with the CoCoNuT code, which implements general relativistic hydrodynamics employing the conformal flatness condition and considering the first 35 ms of the collapse for four progenitor models of masses ranging from 12 to 40 $M_{\odot}$. All of them with solar metallicity. More recently, Abylkairov et. al. (2025) generated two sets of 452 waveforms \cite{abylkairov2025evaluating}, also using the CoCoNuT code. For the first set they used general relativity with the conformal-flatness condition and for the second set, Newtonian hydrodynamics with the general relativity effective potential was implemented. In both cases, the axisymmetry of the core in the early post-bounce oscillations was considered for the extraction of the 2D signal \cite{ott2007rotating}. In these simulations, magnetic fields are not considered, since for the early stages of the collapse, they do not influence the dynamics of the stellar core \cite{mosta2014magnetorotational}. Both catalogs of simulations share four EOS in common: SFHo \cite{steiner2013core}, LS220 \cite{lattimer1991generalized}, HSDD2 \cite{hempel2010statistical} and GShenFSU2.1 \cite{shen2011second}.  We exclude LS220 from our analysis because of the constraints 
 Tews, et. al. (2017) \cite{tews2017symmetry}  presented on the symmetry energy parameters of neutron matter, based on theoretical grounds and on experimental results of cold atoms. Their work provided a summary of the allowed parameter space regions for nuclear EOS. 
 
 The estimation of peak frequency for the core bounce component and the related ringdown was discussed  in \cite{rover2009bayesian} where they performed a principal component analysis in Gaussian colored noise based on LIGO power spectral density \cite{barsotti2018updated}, and discussed the dependence of the bounce and post-bounce signal on the rotational rate $\beta$. In Pastor-Marcos et al. (2024) \cite{pastor2024bayesian}, a bayesian parameter estimation analysis was performed using zero-mean Gaussian colored noise based on Advanced LIGO and Virgo power spectral densities,
along with a master template. Using approximately 400 signals from the Richers catalog, they aligned them to the time of the bounce which allowed them to compare the signals, then normalized the amplitude by a factor of  $D \Delta h_{+}$, and rescaled by using $f_{\text{peak}} \propto \rho_c$, the peak frequency of the signal, giving a normalized waveform which is the master template. The physical parameters estimated were: the polarization angle $\psi$, peak frequency in post-bounce oscillations $f_{\text{peak}}$, the difference between the largest peak of the signal $h_{+}(t)$ and the smallest peak, this is, $\Delta h_{+} = \max(h_1,h_3) - h_2$. This latter parameter combined in the expression: $D\Delta h_{+}\sin^2 \iota$ with $D$ the distance, $\iota$ the angle between the core's axis of rotation and the line of sight of the observer. They reported that for a galactic distance of around 10 kpc, it is possible to recover the peak frequency and amplitude with an accuracy better than 10\% for $\sim$ 80\% and $\sim$ 60\% of the signals, respectively \cite{pastor2024bayesian}. Also, these parameters enabled the possibility to recover the physical ones such as $\beta$ with an accuracy of $\sim 25\%$ , having a larger uncertainty in the estimation when using waveforms different from the Richers ones.

The first phenomenological model for the core bounce component was introduced in Villegas et al. (2024)\cite{laura}. They used an analytical expression to fit the core bounce signal and then, perform PE using maximum likelihood estimation and matched filtering in real interferometric noise from the observing run of O3, injecting the Richers waveforms given in Table \ref{tab:profiles}. Compared to Pastor-Marcos et. al. results, they do not include the early post-bounce oscillations in their analysis and also give a theoretical minimum for the error, which is consistent with the accuracies reported in \cite{pastor2024bayesian}. Both works assume that the GW is detected. In the project described in the current paper we make the same assumption and focus on a parameter estimation follow up approach.

More recently  Abylkairov. S et al. (2025) \cite{abylkairov2025evaluating} 
produced a new database of core bounce waveforms and compared
different machine learning algorithms, with the goal to classify the EOS (for only one progenitor model) in the absence of noise. They provide an order of magnitude estimate of the impact of noise based on the source distance. They report that for Advanced LIGO A +, the classification accuracy for optimally oriented source achieves $\gtrsim 70 \%$ up to 20 kpc \cite{abylkairov2025assessing}. Those results are consistent with the ones found in this paper, since for only one optimally oriented detector the accuracy at 10 kpc was around 66\%. The value of this and other orientation-related metrics such as accuracy, are expected to decrease by a factor of 2/5 from optimal orientation to an average of all orientations (see appendix \ref{sec:apendice}).

In this paper we expand the analytical model proposed by \cite{laura} to include one more parameter related to the time-scale of the core bounce phase. The benefit of this modification is assessed through fitting factor analysis and bayesian hypothesis testing. 
 We perform PE with a  Markov Chain - Monte Carlo (MCMC) implementation in real interferometric noise. We also assess the robustness of our MCMC implementation in terms of the sensitivity of posterior distributions to different bayesian priors in Gaussian Colored Noise based on LIGO PSD, and in  O3L1 real interferometric noise.

In this study, the range of possible values for the rotational parameter is divided in two parts: In the regime where $0.02<\beta<0.08$ the angular momentum is related to the progenitor and for values of $\beta > 0.08$, where other mechanisms could play a role in the increase of angular momentum. The physics of the progenitors in the two regimes could be different. For example, the explosion mechanism of the blue supergiant progenitor SN1987A could be explained by the capture of a companion star in a common envelope, produced by a mass-transfer phase \cite{podsiadlowski1990merger} . In addition to the previous consideration, in \cite{heger2000presupernova}, the angular momentum distribution for rapidly-rotating progenitors involves the presence of different presupernova and supernova mechanisms such as magnetic fields that play a more important role in the post-bounce dynamics of the collapse and produce gamma-ray bursts. Also, the centrifugal support for high rotation rates, halts the core bounce phase, collapsing directly to a PNS. In this paper we estimate $\beta$ in the whole range but focus on the EOS estimation only for the lower $\beta$ regime.

In section \ref{sec:dos} we describe the databases used in this study. Section \ref{sec:tres} introduces the parameter estimation methodology. In section \ref{sec:cuatro} we present the parametrized analytical model for the core bounce phase, discuss the physical interpretation of its parameters, and introduce the extended model. Model validation and Bayesian model comparison are presented in section \ref{sec:cinco}. Parameter estimation results in realistic interferometric noise and prior-sensitivity analysis are discussed in section \ref{sec:seis}. Section \ref{sec:siete} is dedicated to the implications of the inferred parameters for constraining the nuclear EOS. Finally, section \ref{sec:ocho} summarizes our findings.

\section{Data from numerical simulations}
\label{sec:dos}

One of the catalogs we will use in this paper was published by Richers et. al. (2017) \cite{richers2017equation}. It is made up of around 1800 2D axisymmetric simulations, from which we have taken only a subset of 126 waveforms that undergo collapse and sample the rotational profiles and EOS given in table \ref{tab:profiles}. Also, we used the Abylkairov catalog of 864 waveforms which is divided up into two groups; the first one is the General Relativity group, which includes waveforms that were generated with numerical relativity. The other set is called GREP (General Relativistic Effective Potential), which is an approximation for strong gravitational fields but does not include the full GR effects. We have decided to drop all the GREP waveforms and just focus on the 452 remaining ones.

\begin{table}[t]
\resizebox{0.9\linewidth}{!}{%
\begin{tabular}{|l|l|l|}
\hline
Rotational Profile {[}km{]} &
  $\Omega${[}rad $s^{-1}${]} &
  No. of profiles \\ \hline
A1 (300) &
  4.0 - 12.0 &
  12 \\ \hline
A2 (467)    & 4.0 - 7.0 & 24 \\ \hline
A3 (634)    & 3.0 - 7.0 & 48 \\ \hline
A4 (1268)   & 2.0 - 6.0 & 18 \\ \hline
A5 (10,000) & 2.0 - 4.0 & 24 \\ \hline
\end{tabular}%
}
\caption{\justifying Features considered to select a subset of waveforms from Richers Catalog. Each rotational profile has corresponding values for the angular velocity of rotation ($\Omega$), degree of differential rotation ($A$) and five equations of state: LS220, BHBL, HSDD2, GShenFSU2.1, SFHo.}
\label{tab:profiles}
\end{table}

\begin{figure}[h!]
    \centering
    \includegraphics[width=0.8\linewidth]{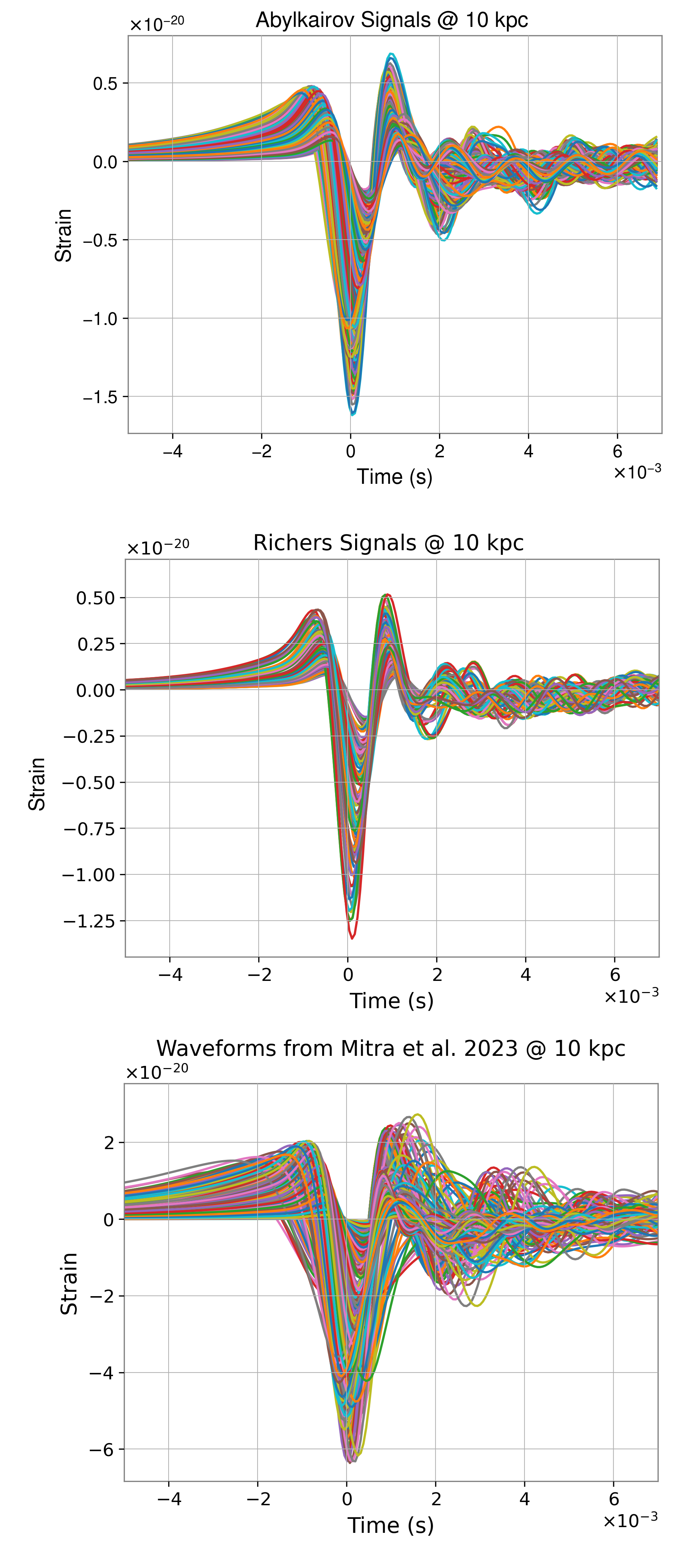}
    \caption{\justifying Set of 126 waveforms from Richers Catalog (bottom) with characteristics from table \ref{tab:profiles} for a $12 M_{\odot}$ progenitor. Set of 452 GR waveforms for a s12 red giant model of about $10.9 M_{\odot}$ from Abylkairov Catalog (top), sampled at a 10 kHz rate.}
    \label{fig:catalogs}
\end{figure}

\begin{figure}
    \centering
    \includegraphics[width=1\linewidth]{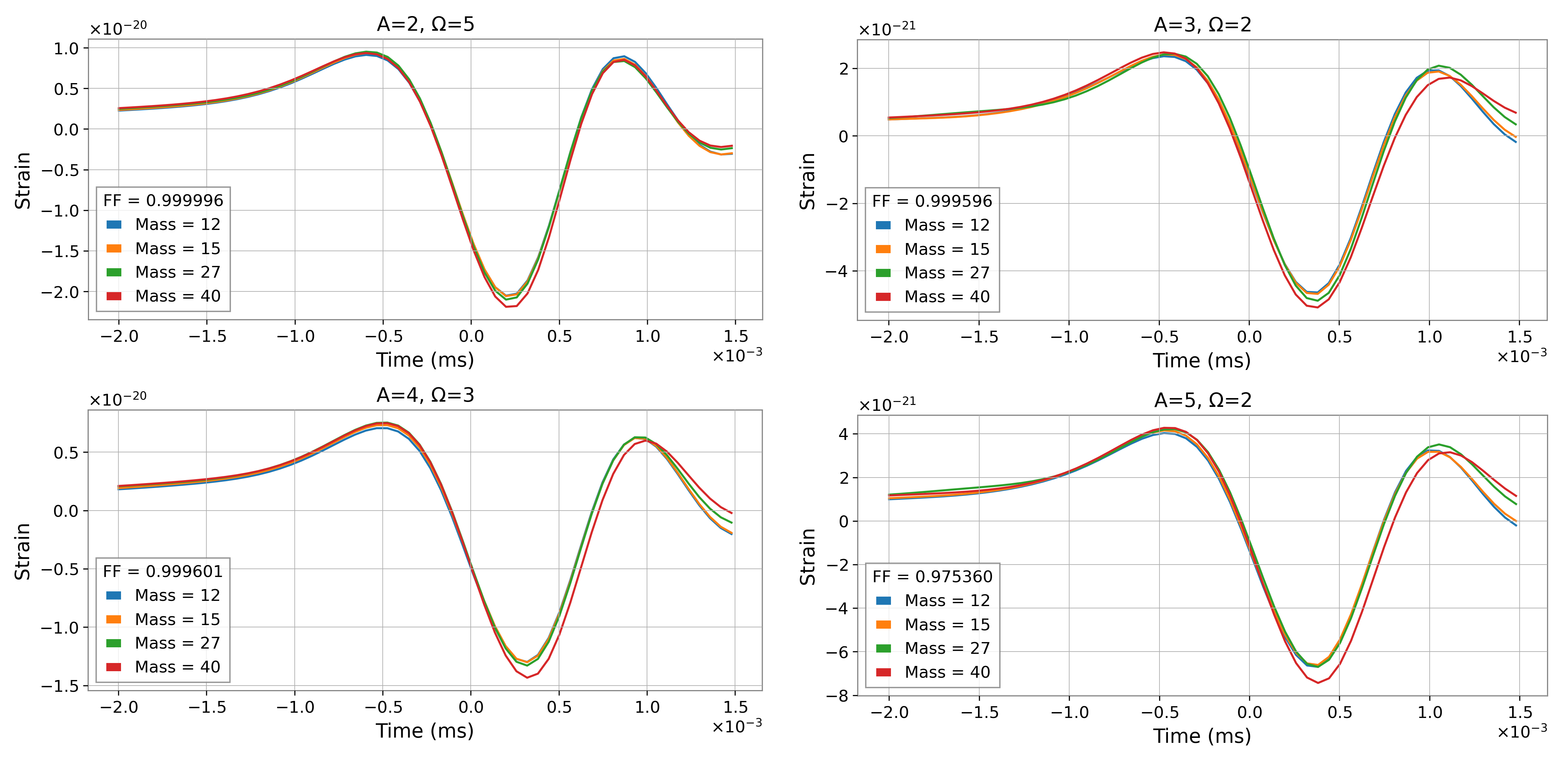}
    \caption{\justifying Waveforms generated for the paper by Mitra, et al. (2023) \cite{mitra2023exploring}. Each panel shows waveforms for different differential rotation $A$ [km] and angular velocity $\Omega$ [rad/s]. The legend compares the fitting factor between the waveforms generated by the highest progenitor mass (40 $M_\odot$) in red and the lowest (12 $M_\odot$) in blue.}
    \label{fig:mitra_masses}
\end{figure}

All of these datasets along with the one used in Mitra, A. et al. (2023) (to which we will refer to as the Mitra catalog for simplicity), show that the morphology of the gravitational wave signal for rapidly rotating progenitors, exhibits three characteristic peaks at the beginning, which correspond to the core bounce, which is present in the vast majority of the signals. 
In figure \ref{fig:catalogs}, we show the signals for the three catalogs, after applying a Butterworth low-pass filter at 800 Hz to remove numerical artifacts as discussed in \cite{laura}. Figure \ref{fig:mitra_masses}, shows the similarity between waveforms of different progenitor masses of 12, 15, 27 and 40 $M_{\odot}$, such that we can state that the core bounce is largely independent of the progenitor mass for the same angular momentum distribution \cite{ott2012correlated,mitra2023exploring}.

\begin{figure}[h]
\centering
\includegraphics[width=1\linewidth]{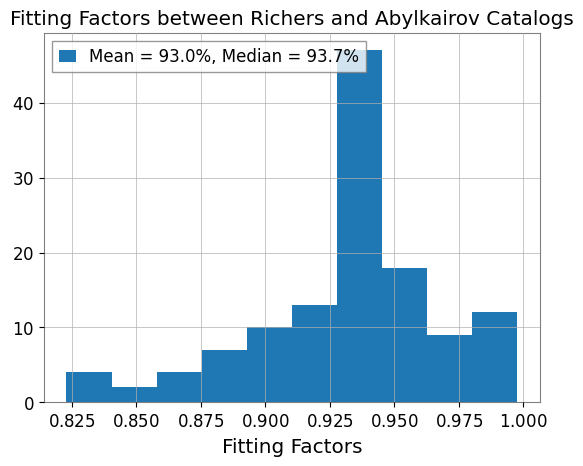}
\caption{\justifying Histogram for the Fitting Factors between signals with same $\beta $ and EOS in Abylkairov catalog to the ones selected from Richers, following the rotational profiles mentioned in table \ref{tab:profiles}}.
\label{fig:FF_AB_RC}
\end{figure}

\begin{figure}
    \centering
    \includegraphics[width=1\linewidth]{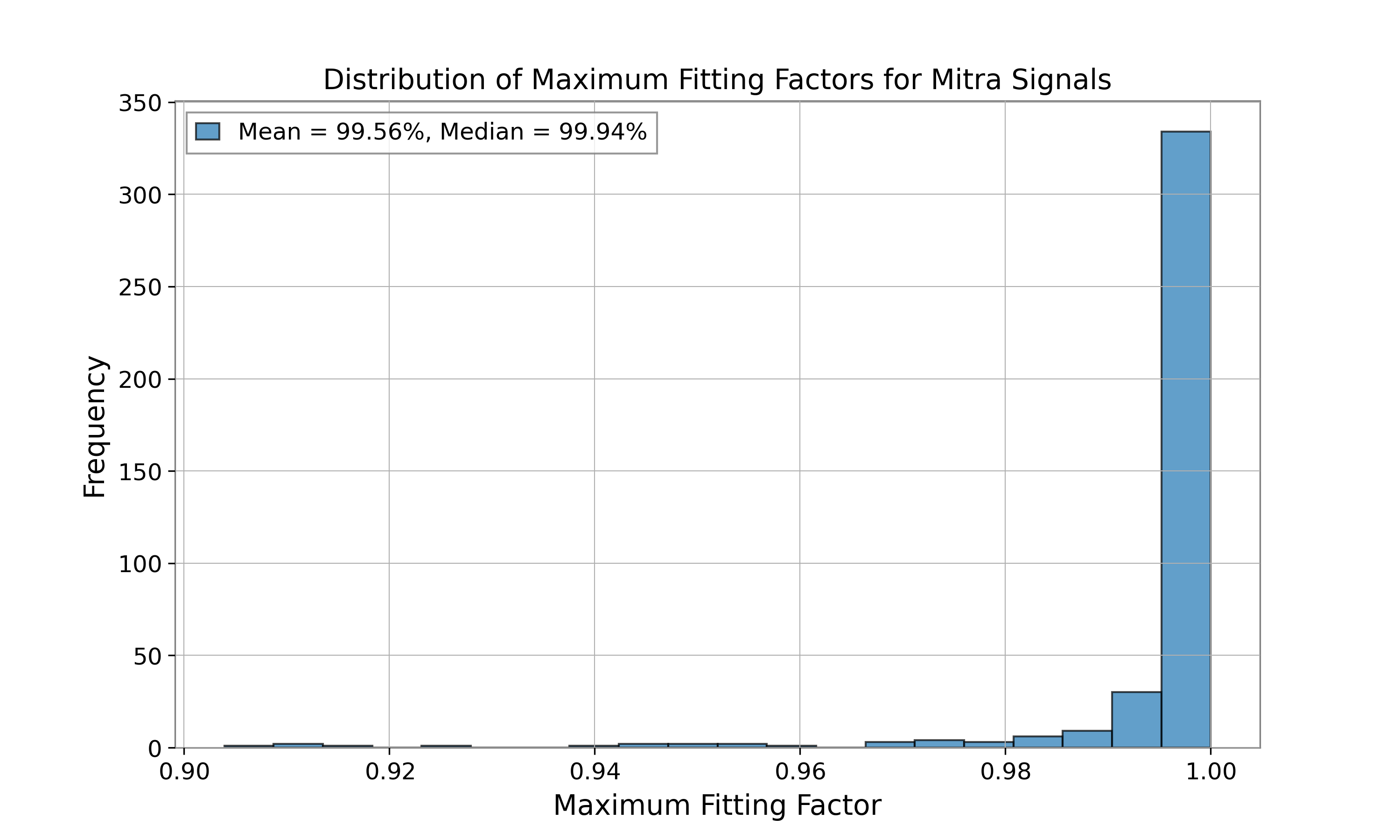}
    \caption{\justifying Histogram of the Maximum Fitting Factors yielded by comparing each waveform from the Mitra Catalog with Abylkairov signals. We compared each signal pair-wise and kept the maximum fitting factor.}
    \label{fig:max_ff_mitra}
\end{figure}

The consistency between Abylkairov catalog and the other two was calculated using the Fitting Factor definition \cite{apostolatos1996construction}. We took the 126 chosen Richers \cite{richers2017equation} signals and the corresponding ones in the Abylkairov catalog for values of $\beta$ within  $2\%$, since both catalogs have not simulated exactly the same values for $\beta$. The resulting fitting factors are shown in figure \ref{fig:FF_AB_RC}. Similarly, each of the signals in Mitra catalog was compared to all the waveforms in Abylkairov and we kept the maximum fitting factor. These are shown in figure \ref{fig:max_ff_mitra}, with a mean of 99.56\% and a median of 99.94\% in O3 noise.

\section{Detection and Parameter Estimation Strategies}
\label{sec:tres}

\subsection{Detector response}
\label{sec:tresA}

For this initial study, we decided to consider the case of a single interferometer, Livingston, whose response $x(t)$ is given by:

\begin{equation}
    \label{eq:det_response}
    x(t) = h(t;  \lambda_j) + n(t).
\end{equation}

The GW strain  is embedded  in real O3aL1 noise \cite{abbott2022ligoO3} $n(t)$, unless stated otherwise. Note that $h(t; \lambda_j)$ depends on a set of physical parameters $\{\lambda_j\}$ and in the long-wave limit \cite{creighton2012gravitational,thorne2000gravitation}  we can expand it in terms of the metric perturbation polarizations $h_{+},h_{\times}$ and the corresponding antenna patterns $F_{+}(\theta,\phi,\psi),F_{\times}(\theta,\phi,\psi)$ in the following way:

\begin{equation}
    \label{eq:ant-pattern}
    h(t, \lambda_j) = F_{+}h_{+}(t, \lambda_j) + F_{\times}h_{\times}(t, \lambda_j).
\end{equation}

As a consequence of symmetry in 2D simulations we took $h_{\times} = 0$, we also took for simplicity an optimal orientation ($F_{+} = 1, F_{\times} = 0$). For a radiating source, we have also taken into consideration the angle between the axis of rotation of the source and the line of sight $\iota$, which is set to be zero, meaning an optimaly-oriented source. We have estimated that the SNR results for a random orientation of the detector and the source is reduced by a factor of 2/5 with respect to the optimal SNR considered in here. In the Appendix, we give the full derivation of this fraction.

When considering noise in the detector to be stationary and gaussian, we can define an inner product $\langle \cdot |\cdot\rangle$ in the space of functions $h(t)$ using the Power Spectral Density $S_n(f)$ and define the Probability Density Function (PDF) of a signal embedded in noise:

\begin{equation}
    \label{eq:PDF}
    p(n) \propto e^{-\langle n|n\rangle/2}.
\end{equation}

There are two main paradigms in the PE problem of GW signals: The frequentist one and the Bayesian one. Both of them will be briefly compared in following subsections.

\subsection{Frequentist Parameter Estimation}

In this approach, it assumes the estimated physical parameters $\lambda_j$ of the signal are deterministic quantities and the detector noise is the only random quantity. Moreover, according to the Neyman-Pearson lemma, under some specific circumstances \cite{scott2005neyman}, the optimal detection statistic is defined as the ratio of the PDFs for the null hypothesis $H_0$ (only noise present in $x(t)$) and the alternative hypothesis $H_1$, where there is indeed a GW signal. The likelihood ration then can be written as:

\begin{equation}
    \label{eq:likeli_ratio}
    \Lambda = \frac{p(x|H_1)}{p(x|H_0)}.
\end{equation}

    This quantity is used for detection using methods like matched filtering \cite{janquart2020gravitational}, but it can also be applied in PE. Taking the PDF from equation \ref{eq:PDF}, we can maximize the log-likelihood $L = \log \Lambda$ and find a given set of $\{\lambda_j^{\text{max}}\}$ which maximizes probability of detection. According to Cutler \& Flannagan (1994) \cite{cutler1994gravitational}, for large Signal-to-Noise Ratio (SNR) values, the covariance matrix in the frequentist case allows to estimate the measurement accuracies $\sqrt{\Delta \lambda_j}$ and it coincides with the diagonal elements of the inverse Fisher Information Matrix. Giving us the Cramér-Rao Lower Bound for error measurement. Higher order corrections are discussed in \cite{vitale2011application}.

    \begin{figure}
        \centering
        \includegraphics[width=1\linewidth]{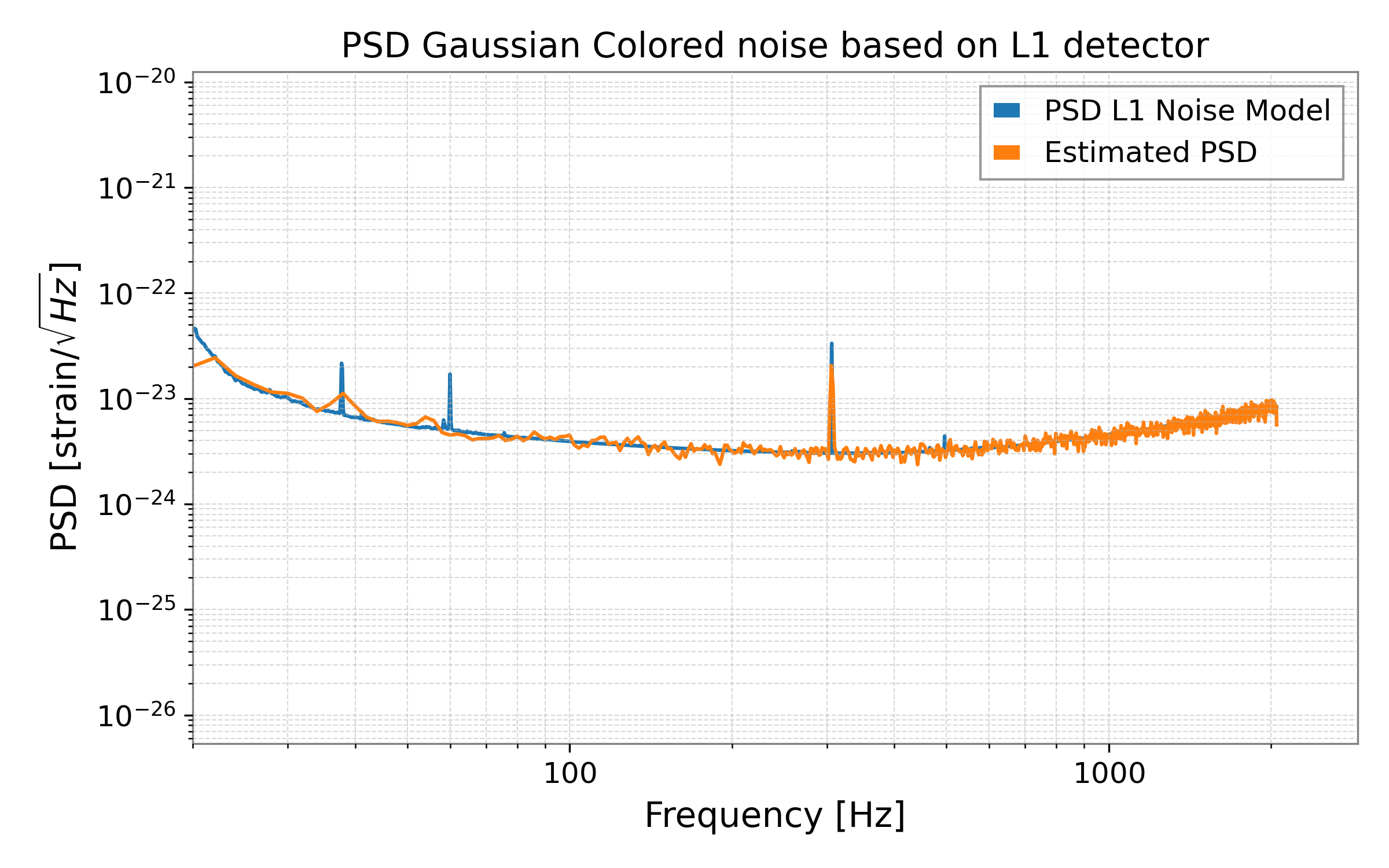}
        \caption{\justifying Curve of the noise model in L1 of Advanced LIGO (blue) \cite{barsotti2018updated} and Estimated PSD (orange).}
        \label{fig:psd_lgcn}
    \end{figure}

\subsection{Bayesian Inference and hypothesis testing}

 The bayesian approach takes the physical parameters as random variables with a probability distribution. This paradigm is already used in GW analysis for both detection and PE \cite{littenberg2009bayesian}, allowing to apply algorithms like MCMC and estimate the posterior PDFs for each parameter using the expected value of the distribution as a quantity which will maximize the posterior PDF. This expected value is called Bayes Estimator \cite{finn1992detection, cutler1994gravitational} or Maximum a Posterior (MAP), in contraposition with Maximum Likelihood Estimator (MLE), and is defined as:

\begin{equation}
    \label{eq:bayes_est}
    \hat{\lambda}_{\text{BE}} = \arg \max_{\lambda} p(\lambda|x).
\end{equation}

The PDF comes from Bayes Theorem:

\begin{equation}
    \label{eq:Bayes_theorem}
    p(\lambda|x) = \frac{\mathcal{L}(x|\lambda)\pi(\lambda)}{Z},
\end{equation}

which includes the likelihood function $\mathcal{L}$, the marginalized evidence $Z$, that is a normalization constant and the prior assumptions about the parameters $\pi (\lambda)$.

In this approach we also calculate measurement accuracy, that can be evaluated from the unbiased Bayes Estimator or Maximum A Posteriori (MAP). The second main difference between frequentist and bayesian approach, is the selection of priors since they impact the mean and variances of the estimated parameters, in the sense that some of them seem to be lower than the Cramér-Rao lower Bound, a theoretical limit \cite{tso2016measuring,chattopadhyay2025prior}. Usually, priors are uninformed if there are no experimental or theoretical constraints that suggest particular distributions for a given set of parameters. It is worth noticing that a uniform prior in a parameter (for example the GW amplitude $h$) is not uniform in any function of such parameter, say $h^2$ \cite{mezzacappa2024gravitational}. 

An additional advantage of the Bayesian framework is that we can use the evidence to calculate the relative probability between different hypothesis (templates in our case), that intend to explain data. Let us say that there is an analytical model $M$ that has $m$ parameters and a given functional form $h_{M}(t; \lambda_i)$ and another model $N$ with $n$ parameters, with its own functional form $h_N{(t;\lambda_j)}$, both of them trying to fit the numerical waveforms of some catalog $C$. Then, we can calculate the Bayes Factor between both models:

\begin{equation}
    \label{eq:bayes_factor}
    \mathcal{B}^{M}_{N} = \frac{Z_M}{Z_N}.
\end{equation}

This is the ratio of the marginalized evidence, defined as:
\begin{equation}
    Z_{M} = \int_{\Omega_M}\mathcal{L}_M(x|\lambda)\pi_{M}(\lambda)d\lambda,
\end{equation}
where the integral is taken all over the parameter space $\Omega_M$ spanned by the physical parameters $\lambda$. The difference in models can be manifested, as we saw, in the functional form, the priors and the number of parameters. The bayes factor is a metric for discriminating between templates and of introducing  changes to them. Nevertheless some bayesian inference algorithms such as MCMC do not calculate the evidence directly, which is needed for the Bayes Factor, so we need other methods to do so, in this study we rely on Nested Sampling \cite{ashton2022nested}.

\section{Core Bounce parametrized analytical model}
\label{sec:cuatro}

\subsection{Functional form of the core bounce model}
\label{sec:cuatroA}

Being the first 3 peaks of the GW core bounce emission from rapidly rotating of CCSNe a deterministic deterministic component, we can adopt a phenomenological analytic model in terms of physical parameters $\lambda_j$. From section \ref{sec:tresA} we assumed an optimal orientation of the detector and source, as well as the symmetry of 2D simulations, so that the strain $h(t) = h_{+}(t,\lambda_j)$.

\begin{widetext}
\begin{equation}
    \label{eq:cbs_eq}
    h_{+}(t;\lambda_j) =\frac{h_1(\beta)}{D}e^{-(t-\mu_1(\tau))^2/2s^2} +  \frac{h_2(\beta)}{D}e^{-(t-\mu_2(\tau))^2/2s^2} + \frac{h_3(\alpha,\beta)}{D}e^{-(t-\mu_3(\tau))^2/2s^2}.
    \end{equation}
\end{widetext}

 This model initially proposed by Villegas, et al. (2025) \cite{laura}, captures the morphology of the core bounce phase from waveforms in Richers catalog. The peaks are three gaussian bells of width $s$ and means $\mu_i$. The quantities $h_i$ are second order polynomials in $\beta$, with constants $C_{ij}$  found using curve fitting on the Richers catalog:
\begin{align}
\label{eq:peaks}
    h_1 &= C_{11} + C_{12}\beta + C_{13}\beta ^2,\\
    h_2 &= C_{21} + C_{22}\beta + C_{23}\beta ^2,\\
    h_{3} &= C_{31} + \alpha\left(\frac{\beta}{0.06}\right)^2,
\end{align}
and the means of the gaussians are:
\begin{align}
    \label{eq:mus}
    \mu_1 &= \tau ,\\
    \mu_2 &= \tau + 0.005 ,\\
    \mu_3 &= \mu_2 + 0.005.
\end{align}

The amplitude of $h_{+}$ depends on the distance $D$ and each of its peaks depend on $\beta$, only the third one has a dependence on $\alpha$, these quantities are related to the overall amplitude of the signal. In the same way, the positions of the peaks are a function only of the parameter $\tau$, the time at which the core bounce occurs, and the constant values of 5 ms summed to each peak adds resemblance of the model to the numerical waveforms. 

The rotational parameter $\beta = T/|V|$ is the ratio between kinetic rotational energy $T$ and binding potential gravitational energy $|V|$. In this work we use values in the interval [0.005,0.18] considering the intersection of values used in numerical simulations from studies such as in Dimmelmeier et al. (2008) Richers et al. (2017) and Abylkairov et al. (2025) \cite{dimmelmeier2008gravitational,richers2017equation,abylkairov2025evaluating}. The time at which the bounce occurs is represented by $\tau$ and is in the interval [-5,7] ms, this latter interval is just the complete duration of the time-window of interest of signals in Abylkairov catalog. The parameter $\alpha$ was introduced to better approximate the amplitude of the third peak, where the EOS role is more important. In the original model \cite{laura} the value of the fourth parameter was fixed to $s = 0.0002$, while here is free to vary.

\section{Template validation and comparison}
\label{sec:cinco}

\subsection{Fitting Factor Analysis}

To validate the decision of introducing a new parameter, we calculated the Fitting Factors given by equation \ref{eq:FF} between the chosen waveforms from Richers catalog and their corresponding signals generated from the analytical model using the estimated parameters from matched filter ($\alpha,\beta,\tau$), then for the same signals in the catalog we generated the corresponding ones using the model but this time using a list of 30 values for the new parameter, where $s \in (1.5\times 10^{-4},3\times 10^{-4})$, we calculated the Fitting Factors for each value in the list and kept the maximum.

\begin{figure*}
    \centering
    \includegraphics[width=1\linewidth]{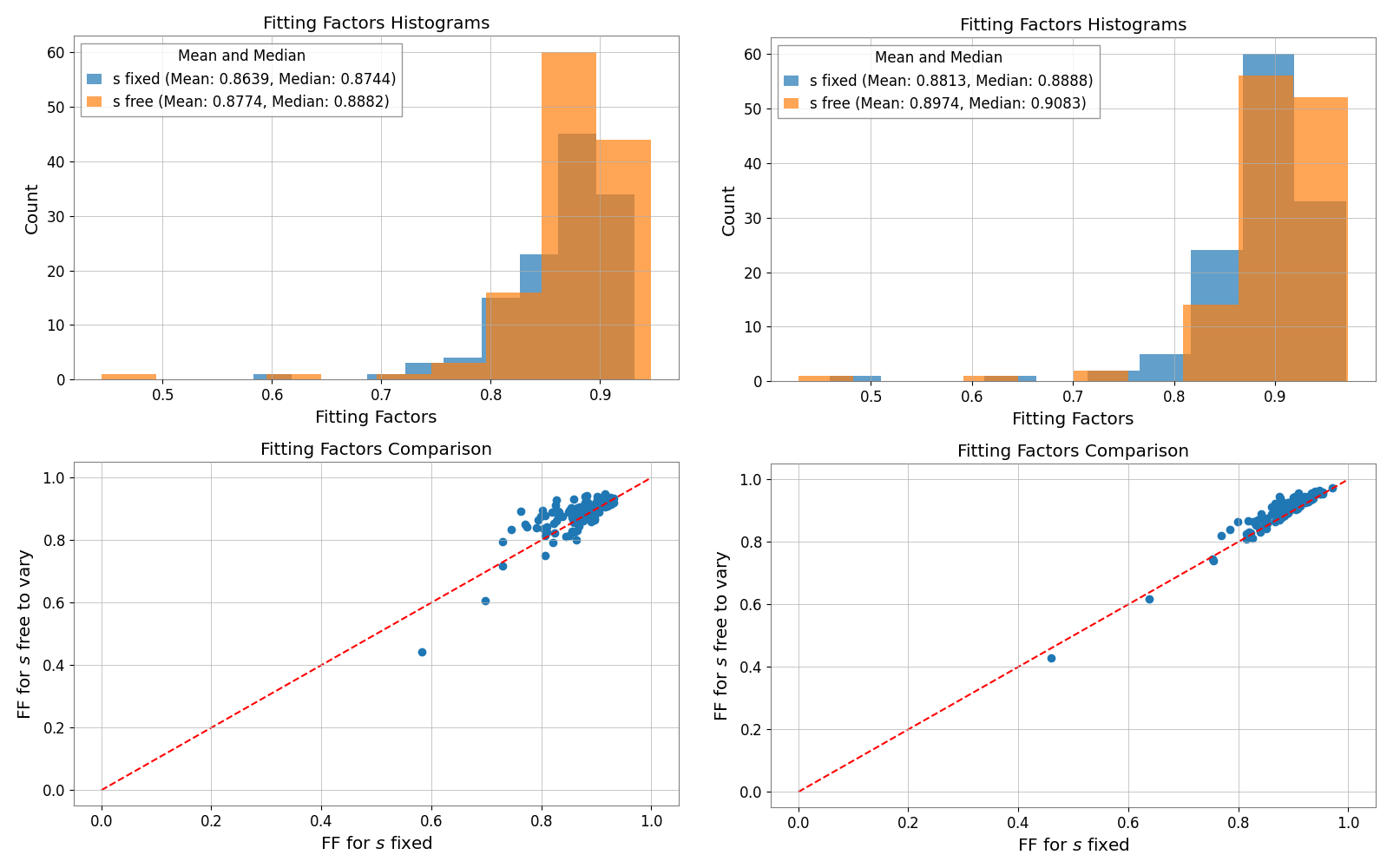}
    \caption{Histogram for the fitting factors of $s$ fixed and $s$ free to vary with their respect mean and median values, including  the post-bounce oscillations (left) and excluding them (right). The bottom panel in the figure compares the fitting factors, considering the case if both were equal (red dashed line).}
    \label{fig:ffs_post}
\end{figure*}

\begin{equation}
    \label{eq:FF}
    FF = \frac{\langle h_s (f) | h_m(\lambda) \rangle}{\sqrt{\langle h_s (f) | h_m(\lambda) \rangle\langle h_s (f) | h_m(\lambda) \rangle}}
\end{equation}.

We see in the histograms of figure \ref{fig:ffs_post} that in the case where we consider $s$ as a free parameter, there is a higher mean and median of the distributions, and the fitting factors improve the way in which the analytical model fits the numerically simulated waveforms from Richers. In the scatter plot, if both fitting factors were equal (and would not is no improvement by letting 's' free to vary) all the dots would lie on the red dashed line, but the grouped dots above the line, indicate a better Fitting Factor when we consider a four-parameter model. In \textbf{Figure \ref{fig:ffs_post}}, we included the first two milliseconds of the post-bounce oscillations, and the resulting fitting factors were reduced by $\approx 1\%$.

\subsection{Bayesian Model comparison}

Another way of validating the decision to add a new parameter comes from the concept of Bayesian Factor related to bayesian decision rule, given by equation \ref{eq:bayes_factor}, it gives the most probable model that better fits data. As was seen earlier, the main difference between MCMC and Nested Sampling is that the latter allows us to calculate the bayesian evidence and it was created specifically for computing this quantity, nevertheless MCMC is more robust than Nested sampling \cite{hogg2018data,thrane2019introduction,ashton2021bilby} in the sense that the samples from the posterior PDF, are more reliable or tend to minimize uncertainty compared to the samples drawn using Nested Sampling. However, is widely known that it can be used to explore a template of varying number of parameters that fit to data, something that is not a direct necessity in MCMC because it does not require to compute marginalized evidence $Z$. For our purposes, we can use the analytical model defined by equation \ref{eq:cbs_eq} as our base model (or bayesian hypothesis) with varying number of parameters. The marginalized evidence for model comparison was calculated further using Nested Sampling algorithm the Bayes Factor defined previously in equation \ref{eq:bayes_factor}.

\begin{figure*}
    \centering
    \includegraphics[width=1\linewidth]{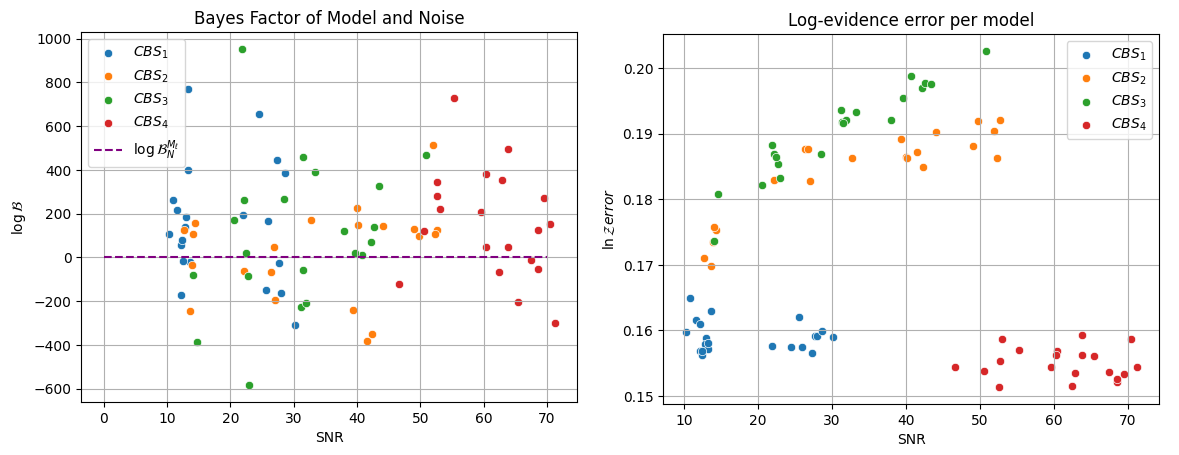}
    \caption{\justifying (Left)Base 10 Log of Bayes factors for each model with different number of parameters (color code) and only noise. The horizontal line represents a threshold, and the estimations above are those identified as more distinguishable from LIGO colored noise. (Right) Error for the natural log of the evidence per model. Each color represents a model with different number of parameters $CBS_i$ for $i \in {1,2,3,4}$. The 1 parametric and 4 parametric models show the smaller errors for the log-evidence error.}
    \label{fig:SNR}
\end{figure*}

For comparing between the same template with varying number of parameters
, we used 80 waveforms from Richers catalog, from different rotational profiles according to the guidelines described in section \ref{sec:dos}. Each Core-Bounce Signal Model is named using the notation $CBS_k$, where $k$ indicates the number of parameters in the model. For $CBS_1$ the parameter considered was $\beta$; for $CBS_2$ $\beta$ and $\tau$, $CBS_3$ $\beta,\tau, \alpha$ and for $CBS_4$ $\beta,\tau,\alpha,s$. On the other hand, figure \ref{fig:SNR} shows the results of each estimation in groups of 20 waveforms, each of these groups corresponding to a model of $n$ parameters for $n \in \{1,2,3,4\}$.  For example, Group 1 consists of 20 waveforms chosen randomly from the set of 126, and this group was used to perform parameter estimation with the $CBS_1$ model, which stands for a 1-parameter Core-Bounce Signal model. According to the Jeffreys' scale \cite{jeffreys1998theory}, if  $1 < \log _{10} \mathcal{B} < 2$ then the evidence against the model in the numerator is strong, and decisive if the value is greater than 2. In figure \ref{fig:SNR}, we compare the model evidence for a different number of parameters against pure noise, which means that the higher the Bayes factors, the more distinguishable is the signal from noise. Also it depicts the natural logarithm of the error for the evidence, which is minimum in the 4 and the 1-parametric models in their corresponding range of signal-to-noise ratio (SNR). From the results of the Bayes factor for each combination of models shown in table \ref{tab:bayes_f} the 1-parametric model is preferred among all the rest, but there is not a strong evidence in its favor compared to the 4-parametric model, so from Bayesian hypothesis testing, we can also state that a four parameter model ($\alpha,\beta,\tau,s$) is preferred to adjust it to the Richers catalog, also we draw this conclusion supported by the previous Fitting Factor analysis and the physically-informed parameters.

\begin{table}[]
\begin{tabular}{|c|c|c|c|c|c|}
\hline
$B^{1}_{2}$ & $B^{1}_{3}$ & $B^{1}_{4}$ & $B^{2}_{3}$ & $B^{2}_{4}$ & $B^{3}_{4}$ \\ \hline
127.60      & 55.32       & 1.44        & -72.28      & -126.16     & -53.89      \\ \hline
\end{tabular}
\caption{\justifying Table of Log base 10 of Bayes factors computed by the different combinations between the models with different number of parameters. For example, $B^{1}_{2}$ stands for the Log base 10 Bayes factor of the 1 parameter ($\beta$) model $CBS_1$; against the 2 parameter ($\beta, \tau$) model $CBS_2$ and the same goes for the rest of the factors. To select the model with a best fit we have to compare between pairs following Jeffrey's scale.}
\label{tab:bayes_f}
\end{table}

\section{Parameter Estimation Results}
\label{sec:seis}

\subsection{Prior Sensitivity Analysis}

\begin{figure}[h]
    \centering    \includegraphics[width=1\linewidth]{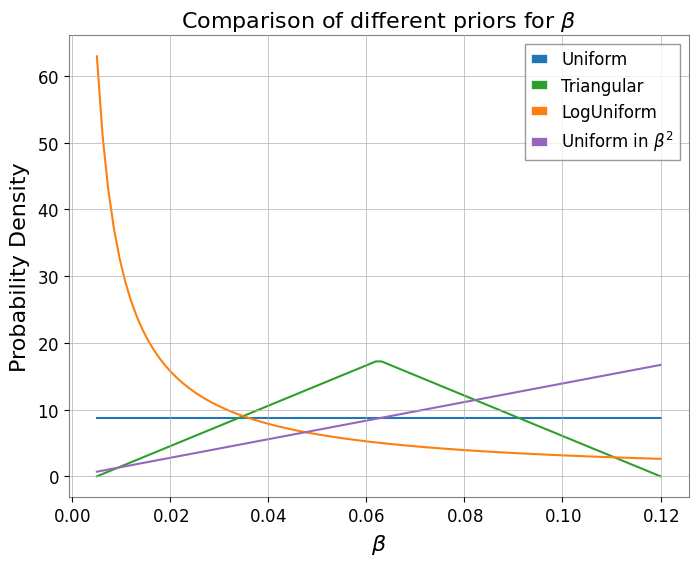}
    \caption{\justifying Prior probability density functions to test prior-sensitivity for $\beta$ parameter. From top to bottom: Uniform, Log-Uniform, Triangular and Uniform in $\beta ^2$. All of them defined in the range $[0.005,0.12]$ and for the triangular prior, the mode was chosen to be the average of the interval boundaries.}
    \label{fig:priors}
\end{figure}

\begin{figure*}
    \centering
    \includegraphics[width=1\linewidth]{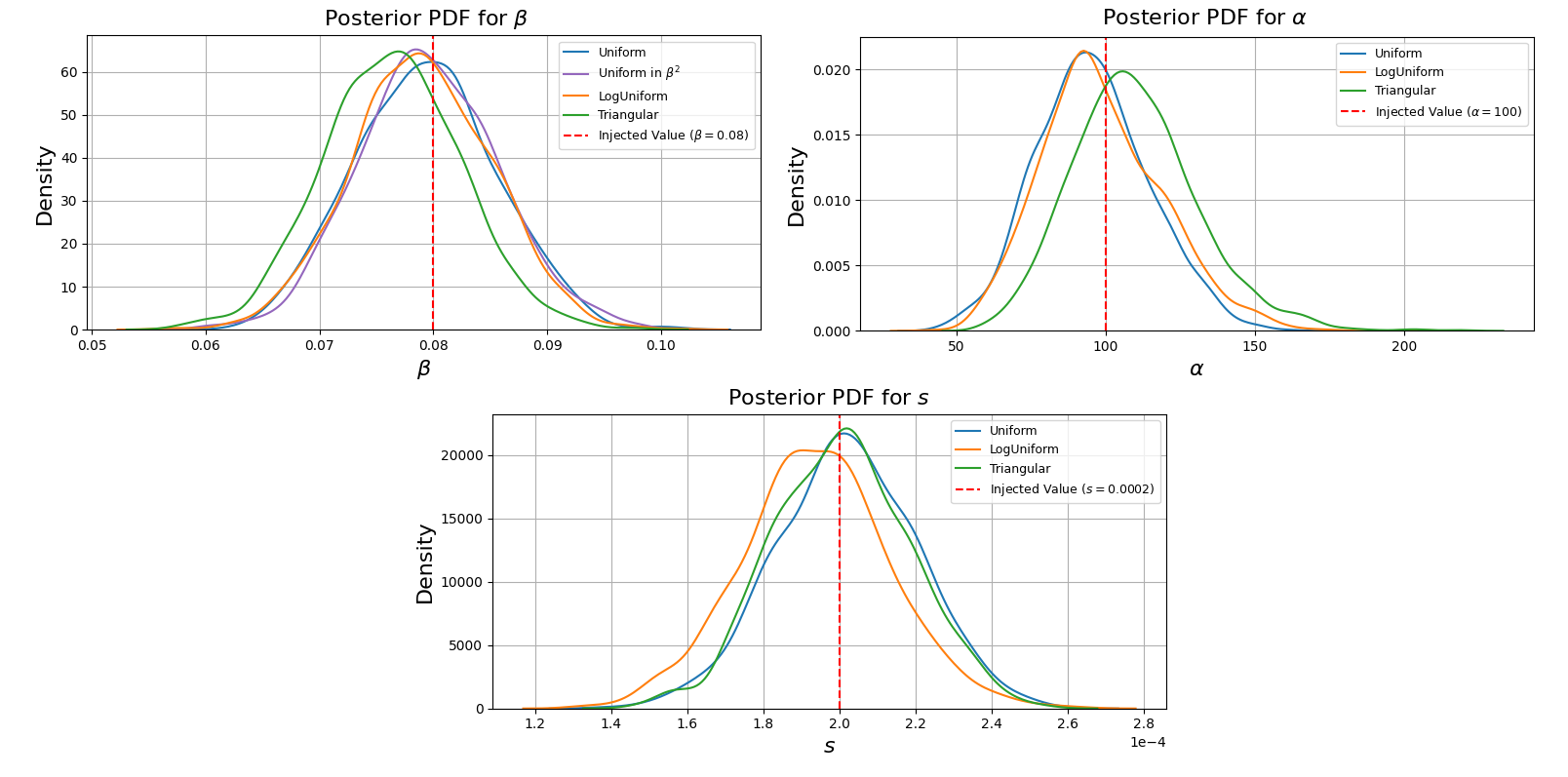}
    \caption{\justifying Posterior PDFs at 10 kpc for 1000 samples in the MC-MC run. In LIGO Gaussian colored noise. Injected values are the red dashed lines. Corresponding to $\beta = 0.08$, $\alpha = 100$ and $s = 2\times 10^{-4}$.}
    \label{fig:KDE}
\end{figure*}

In order to test sensitivity to bayesian priors of the PE estimation process, we decided to use the following priors: Uniform, Triangular, Log-Uniform for three parameters $\alpha,\beta, s$ and a separate case of considering a uniform distribution in the energy of the signal $\beta ^2$, because a uniform in $\beta$ is not equivalent to uniform in $\beta^2$. We took $n = 1000$ samples for the MCMC run with a stopping criteria using the ACT (Autocorrelation Time). This is, to stop after 1000 independent samples from the posterior were drawn. To quantify bias induced by the difference between numerical waveforms and the analytical model, we used the latter one to generate a signal with parameters $\beta = 0.08, \alpha = 100, s = 0.0002$. These results are produced for LIGO Gaussian Colored Noise (LGCN) and O3L1 noise using three different rotational regimes $(0.025, 0.08, 0.11)$ and one signal generated with the templated model all of them at 5 and 10 kpc. After setting the values for priors shown in figure \ref{fig:priors}, we obtained the posterior probability distributions shown in figure \ref{fig:KDE}, the variability of the mean value for each of the priors is find within an error of $10^{-2}$ in gaussian simulated noise at a distance of 10 kpc, and in all the three cases we can see that the triangular prior tends to introduce bias to the mean value of the posterior distributions. This shows that there is indeed a sensitivity that tends to bias the results towards the regions of higher probability of the parameters, indicated by the priors. The Log-Uniform prior has a negative bias in the $\alpha$ parameter, since by definition the smaller values are the most probable to be present in the posterior PDF, the same reasoning can explain the positive bias for the Triangular prior, since for this we need to give the mode, defined in this study as the mean value of the upper and lower bounds for each parameter.

\begin{figure*}
    \centering
    \includegraphics[width=0.8\linewidth]{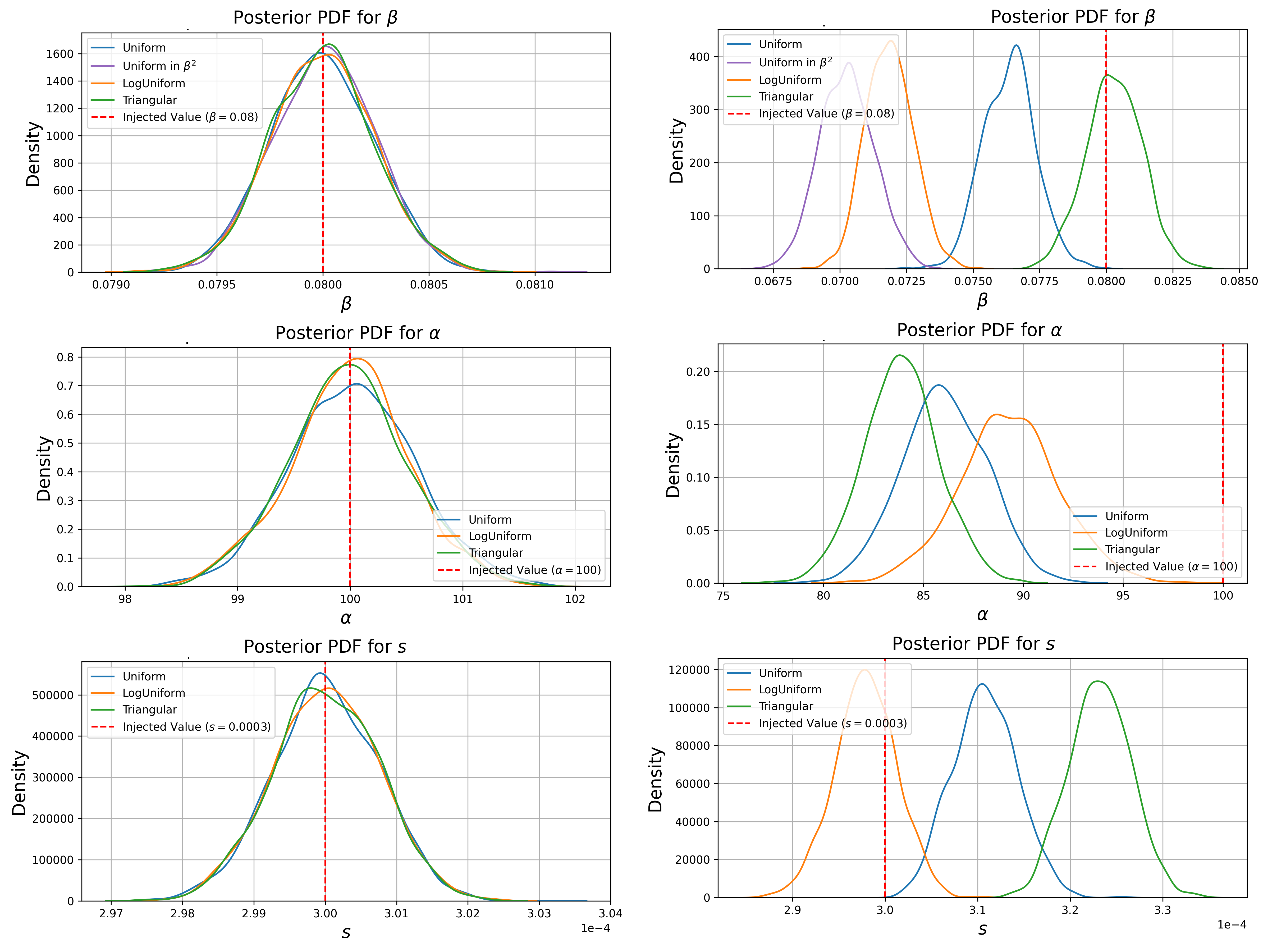}
    \caption{\justifying Posterior PDFs at 10 kpc for 1000 samples in the MC-MC run. Injected values are the red dashed lines. Corresponding to $\beta = 0.08$, $\alpha = 100$ and $s = 3\times 10^{-4}$. Left panel is the noiseless case and the other one the O3 noise case.}
    \label{fig:O3KDE}
\end{figure*}

On the other hand, in  table \ref{tab:two_betas}, we estimated values for the slow and high rotation regimes. The variances for high $\beta$
values are slightly higher compared to the smaller ones, but the order of magnitude stays at $10^{-3}$ in real noise at 10 kpc. Also, the log-uniform prior showed smaller variance.

\begin{table}[h]
\renewcommand{\arraystretch}{1.3}
\setlength{\tabcolsep}{10pt}
\centering
\begin{tabular}{|c|c|c|c|}
\hline
Prior & $\beta\ (10^{-2})$ & $\hat{\beta}\ (10^{-2})$ & $\sigma _{\beta}\ (10^{-3})$ \\ \hline
\multirow{2}{*}{Uniform} 
  & 2.10 & 2.02 & 4.58 \\ \cline{2-4} 
  & 8.90 & 8.99 & 5.71 \\ \hline
\multirow{2}{*}{Log-Uniform} 
  & 2.10 & 1.86 & 4.41 \\ \cline{2-4} 
  & 8.90 & 8.63 & 5.92 \\ \hline
\multirow{2}{*}{Triangular} 
  & 2.10 & 2.45 & 4.56 \\ \cline{2-4} 
  & 8.90 & 8.86 & 5.69 \\ \hline
\end{tabular}
\caption{\justifying This table shows the estimated values with MCMC $\hat{\beta}$ in O3L1 Noise and the corresponding Frequentist Matched Filtered $\beta$. The last column shows the error $\sigma _\beta$ with three different priors.}
\label{tab:two_betas}
\end{table}

\begin{table}[H]
\centering
\begin{tabular}{|ccc|ccc|}
\hline
\multicolumn{3}{|c|}{Uniform $\beta$} & \multicolumn{3}{c|}{Uniform $\beta ^2$} \\ \hline
\multicolumn{1}{|c|}{\begin{tabular}[c]{@{}c@{}}MF \\ $\beta$\end{tabular}} & \multicolumn{1}{c|}{MCMC $\hat{\beta}$} & $\sigma _{\beta}$ & \multicolumn{1}{c|}{\begin{tabular}[c]{@{}c@{}}MF \\ $\beta$\end{tabular}} & \multicolumn{1}{c|}{MCMC $\hat{\beta}$} & $\sigma _{\beta}$ \\ \hline
\multicolumn{1}{|c|}{0.106} & \multicolumn{1}{c|}{9.49e-2} & 2.03e-3 & \multicolumn{1}{c|}{0.106} & \multicolumn{1}{c|}{1.13e-1} & 1.73e-2 \\ \hline
\multicolumn{1}{|c|}{0.083} & \multicolumn{1}{c|}{6.91e-2} & 4.33e-4 & \multicolumn{1}{c|}{0.083} & \multicolumn{1}{c|}{8.18e-2} & 4.32e-3 \\ \hline
\multicolumn{1}{|c|}{0.021} & \multicolumn{1}{c|}{2.76e-2} & 1.74e-4 & \multicolumn{1}{c|}{0.021} & \multicolumn{1}{c|}{2.45e-2} & 2.61e-3 \\ \hline
\end{tabular}
\caption{\justifying PE for three different rotation rates (low 0.025, medium 0.08, high 0.11). Injected values were the estimated ones using the Frequentist Matched Filtering (MF) method and $\hat{\beta}$ are estimated values with MC-MC. These are injections in O3L1 Noise.} 
\end{table}

\begin{table}[ht]
\centering
\label{tab:bias_min_max}
\begin{tabular}{lcccc}
\hline
Parameter & Distance & Min. bias (\%) & Max. bias (\%) \\
\hline
$\beta$ (0.08) & 10 kpc & 0.85 (Uniform) & 4.61 (Triangular) \\
$\beta$ (0.08) & 5 kpc  & 0.46 (Log-uniform) & 10.55 (Triangular) \\
$\alpha$ (100) & 10 kpc & 1.26 (Log-uniform) & 9.66 (Triangular) \\
$\alpha$ (100) & 5 kpc  & 1.19 (Uniform) & 3.31 (Triangular) \\
$s$ (0.0002) & 10 kpc & 0.26 (Triangular) & 3.18 (Log-uniform) \\
$s$ (0.0002) & 5 kpc  & 0.095 (Triangular) & 0.70 (Uniform) \\
\hline
\end{tabular}
\caption{\justifying Minimum and maximum relative bias (in percent) for each parameter
at 5 and 10 kpc in Gaussian LIGO colored noise. The prior producing each extreme value is indicated
in parentheses.}
\end{table}

\begin{table}[h]
\centering
\begin{tabular}{lcc}
\toprule
Parameter & Min. bias (\%) & Max. bias (\%) \\
\midrule
$\beta$ (0.08) & 0.6 (Triangular) & 11.9 (Uniform in $\beta^2$) \\
$\alpha$ (100) & 11 (LogUniform) & 16 (Triangular) \\
$s$ (0.0002) & 49 (LogUniform) & 61 (Triangular) \\
\bottomrule
\end{tabular}
\caption{\justifying O3 noise approximate prior-induced bias estimated from the posterior PDFs for a source distance of 10 kpc. The minimum and maximum biases correspond to the priors that produce the smallest and largest deviation from the injected parameter values.}
\label{tab:prior_bias_summary}
\end{table}

\begin{table*}
\begin{tabular}{|cccccccccc|}
\hline
\multicolumn{10}{|c|}{Summary Statistics at 10 kpc}                                                    \\ \hline
\multicolumn{1}{|c|}{Prior}       & \multicolumn{1}{c|}{Bias ($\beta - \hat{\beta}$)} & \multicolumn{1}{c|}{$\sigma_{\beta}$} & \multicolumn{1}{c|}{$RMSE_{\beta}$} & \multicolumn{1}{c|}{Bias ($\alpha - \hat{\alpha}$)} & \multicolumn{1}{c|}{$\sigma_{\alpha}$} & \multicolumn{1}{c|}{$RMSE_{s}$} & \multicolumn{1}{c|}{Bias ($s - \hat{s}$)} & \multicolumn{1}{c|}{$\sigma_{s}$} & $RMSE_{s}$ \\ \hline
\multicolumn{1}{|c|}{Uniform}     & \multicolumn{1}{c|}{6.810e-04}                    & \multicolumn{1}{c|}{6.178e-03}        & \multicolumn{1}{c|}{4.171e-03}      & \multicolumn{1}{c|}{4.453e+00}                      & \multicolumn{1}{c|}{1.871e+01}         & \multicolumn{1}{c|}{2.551e+01}  & \multicolumn{1}{c|}{-1.874e-06}            & \multicolumn{1}{c|}{1.860e-05}    & 8.832e-06  \\ \hline
\multicolumn{1}{|c|}{Log Uniform} & \multicolumn{1}{c|}{1.017e-03}                    & \multicolumn{1}{c|}{6.109e-03}        & \multicolumn{1}{c|}{3.659e-03}      & \multicolumn{1}{c|}{1.256e+00}                      & \multicolumn{1}{c|}{2.035e+01}         & \multicolumn{1}{c|}{1.583e+01}  & \multicolumn{1}{c|}{6.363e-06}            & \multicolumn{1}{c|}{1.940e-05}    & 1.710e-05  \\ \hline
\multicolumn{1}{|c|}{Triangular}  & \multicolumn{1}{c|}{3.691-03}                    & \multicolumn{1}{c|}{6.017e-03}        & \multicolumn{1}{c|}{7.653e-03}      & \multicolumn{1}{c|}{-9.659e+00}                      & \multicolumn{1}{c|}{2.103e+01}         & \multicolumn{1}{c|}{2.193e+01}  & \multicolumn{1}{c|}{-5.169e-07}            & \multicolumn{1}{c|}{1.800e-05}    & 1.429e-05  \\ \hline
\multicolumn{10}{|c|}{Summary Statistics at 5 kpc}                                                                                                                 \\ \hline
\multicolumn{1}{|c|}{Uniform}     & \multicolumn{1}{c|}{5.348e-04}                    & \multicolumn{1}{c|}{5.655e-03}        & \multicolumn{1}{c|}{1.451e-03}      & \multicolumn{1}{c|}{1.192e+00}                      & \multicolumn{1}{c|}{1.211e+01}         & \multicolumn{1}{c|}{1.104e+01}  & \multicolumn{1}{c|}{-1.391e-06}            & \multicolumn{1}{c|}{8.598e-06}    & 2.421e-06  \\ \hline
\multicolumn{1}{|c|}{Log Uniform} & \multicolumn{1}{c|}{3.691e-04}                    & \multicolumn{1}{c|}{3.630e-03}        & \multicolumn{1}{c|}{3.659e-03}      & \multicolumn{1}{c|}{1.433e+00}                      & \multicolumn{1}{c|}{1.982e+01}         & \multicolumn{1}{c|}{1.547e+01}  & \multicolumn{1}{c|}{2.393e-07}            & \multicolumn{1}{c|}{3.540e-06}    & 1.778e-06  \\ \hline
\multicolumn{1}{|c|}{Triangular}  & \multicolumn{1}{c|}{8.443e-03}                    & \multicolumn{1}{c|}{6.892e-03}        & \multicolumn{1}{c|}{7.413e-02}      & \multicolumn{1}{c|}{-3.305e+00}                      & \multicolumn{1}{c|}{2.753e+01}         & \multicolumn{1}{c|}{2.124e+01}  & \multicolumn{1}{c|}{1.897e-07}            & \multicolumn{1}{c|}{8.714e-06}    & 1.851e-05  \\ \hline
\end{tabular}
\captionof{table}{Summary statistics for prior sensitivity at 10 kpc (top) and at 5 kpc (bottom) in LIGO Gaussian Colored Noise.}

\end{table*}

\subsection{Parameter Estimation in O3 noise}

\begin{figure*}
    \centering
    \includegraphics[width=1\linewidth]{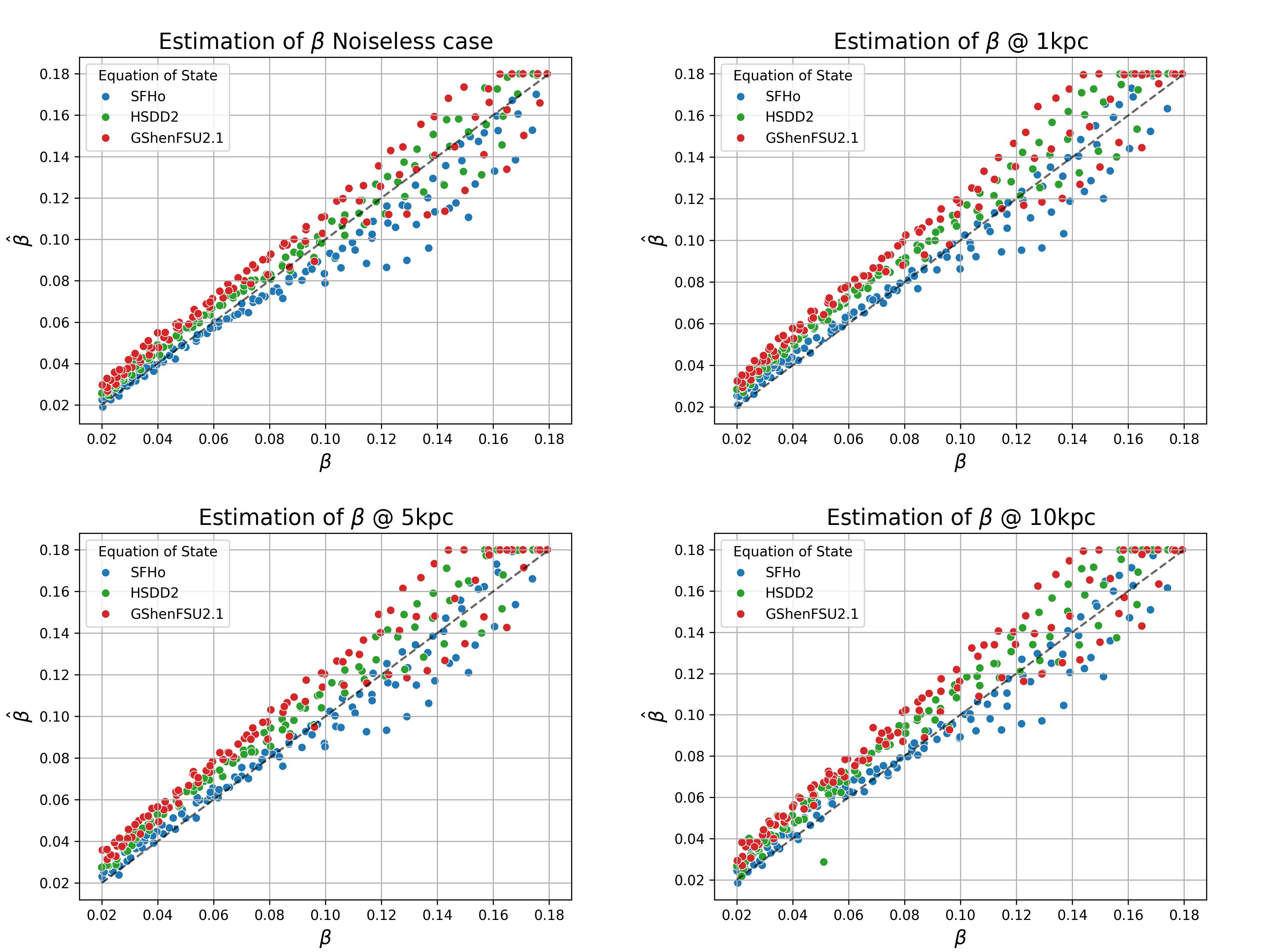}
    \caption{\justifying Estimated value $\hat{\beta}$ vs. $\beta$ at bounce at 5 kpc (left) and 10 kpc (right). The black dashed line is $\hat{\beta} = \beta$}.
    \label{fig:beta_vs_betahat}
\end{figure*}

We used O3L1 noise and the Abylkairov catalog. This project made use of the \texttt{Bilby}'s MCMC sampler, drawing 1,000 samples using a  whitened time-series along with a band-pass filter to reduce noise-related bias in the PE results. In figure \ref{fig:beta_vs_betahat} the results of PE for the noiseless case, 1, 5 and 10 kpc are shown. We observe that the estimated values are more accurate for LS220 and SFHo, while HSDD2 and GShenFSU2.1 show a systematic over-estimation compared to the value of $\beta$ at bounce. Nevertheless, estimated values follow a trend around the black dashed line, that implies an estimation that would have had a 100\% of accuracy.

\section{Estimation of Equation of State}
\label{sec:siete}
\begin{figure*}
    \centering
    \includegraphics[width=0.8\linewidth]{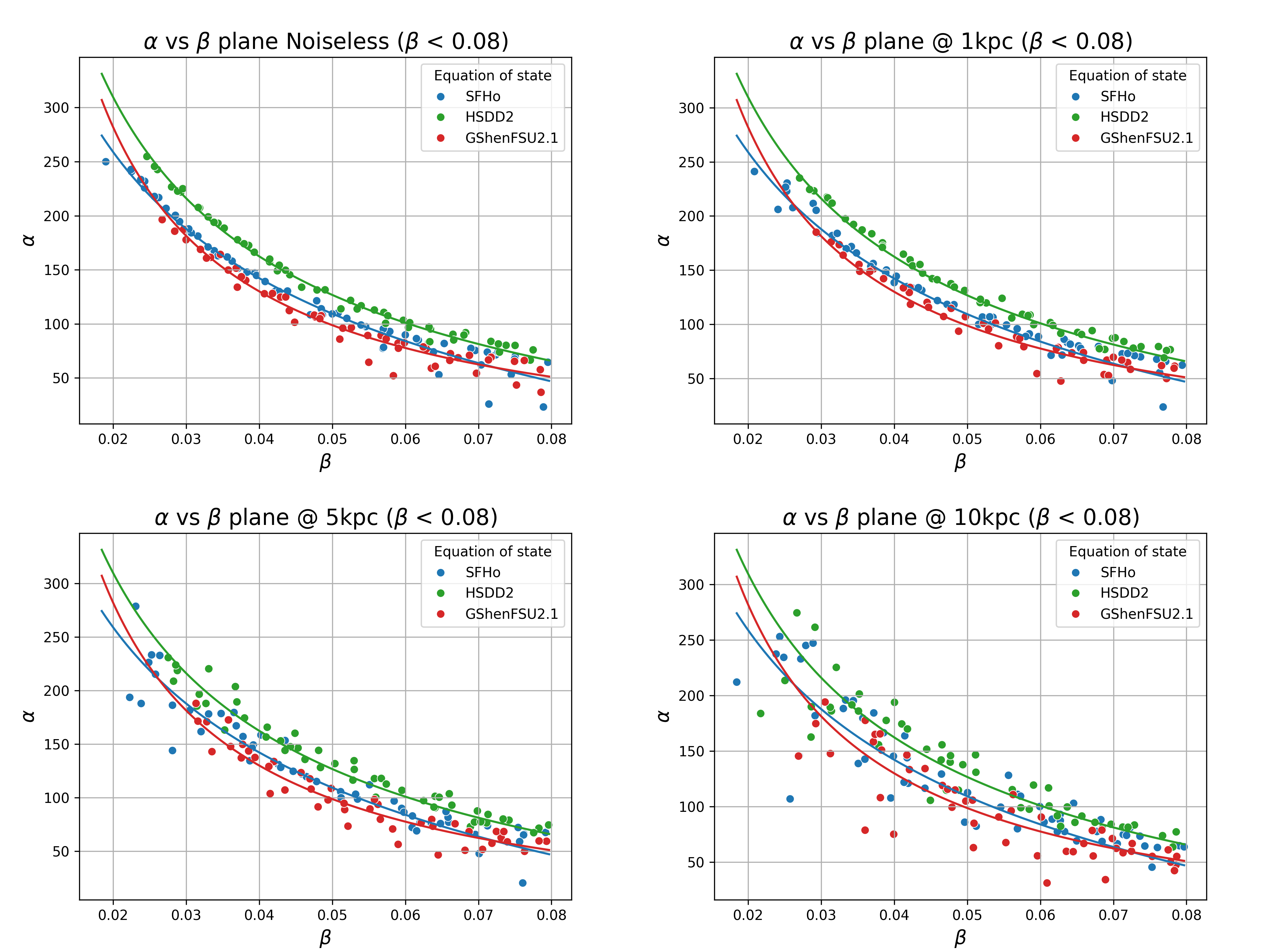}
    \caption{Fit of the Power Law $\alpha(\beta) = A\beta^{B} + C$, for the Noiseless case, 1, 5 and 10 kpc.}
    \label{fig:Power-Lawfits}
\end{figure*}

As stated in previous sections, the parameter $\alpha$ is related to the amplitude of the third peak, as described in equation \ref{eq:peaks}. We propose a power law relationship between the estimated $\alpha$ and $\beta$ as a curve to discriminate between different EOS. In figure \ref{fig:Power-Lawfits} the trend is more visible for values of beta smaller than 0.08 where we assessed the capability to estiate the EOS.
We decided to keep this $\beta$ upper limit in determining a Power-Law curve $A\beta^{B} + C$ for each EOS. The estimated coefficients for each curve are given in table \ref{tab:PL_EOS}. 

\begin{table}[htbp]
\centering
\caption{Coefficients of the Power-Law for each EOS.}
\label{tab:PL_EOS}
\begin{tabular}{lrrr}
\hline
\textbf{EOS} & \textbf{A} & \textbf{B} & \textbf{C} \\
\hline
SFHo        & 126.646829 & -0.364988 & -268.029746 \\
HSDD2       &  36.486333 & -0.623400 & -108.593150 \\
GShenFSU2.1 &  18.431847 & -0.747374 &  -73.770157 \\
\hline
\end{tabular}
\end{table}

The curve in the $\alpha-\beta$ plane has a decreasing trend. For larger values of $\beta$ there is a stronger centrifugal support that would hinder the core from reaching the high densities where rebound effects, like the amplitude of the third peak. 
This is also consistent with the interpretation that the bounce for small rotation rates is pressure-dominated and the EOS plays a more important role during this phase.

\begin{figure*}
    \centering
    \includegraphics[width=0.8\linewidth]{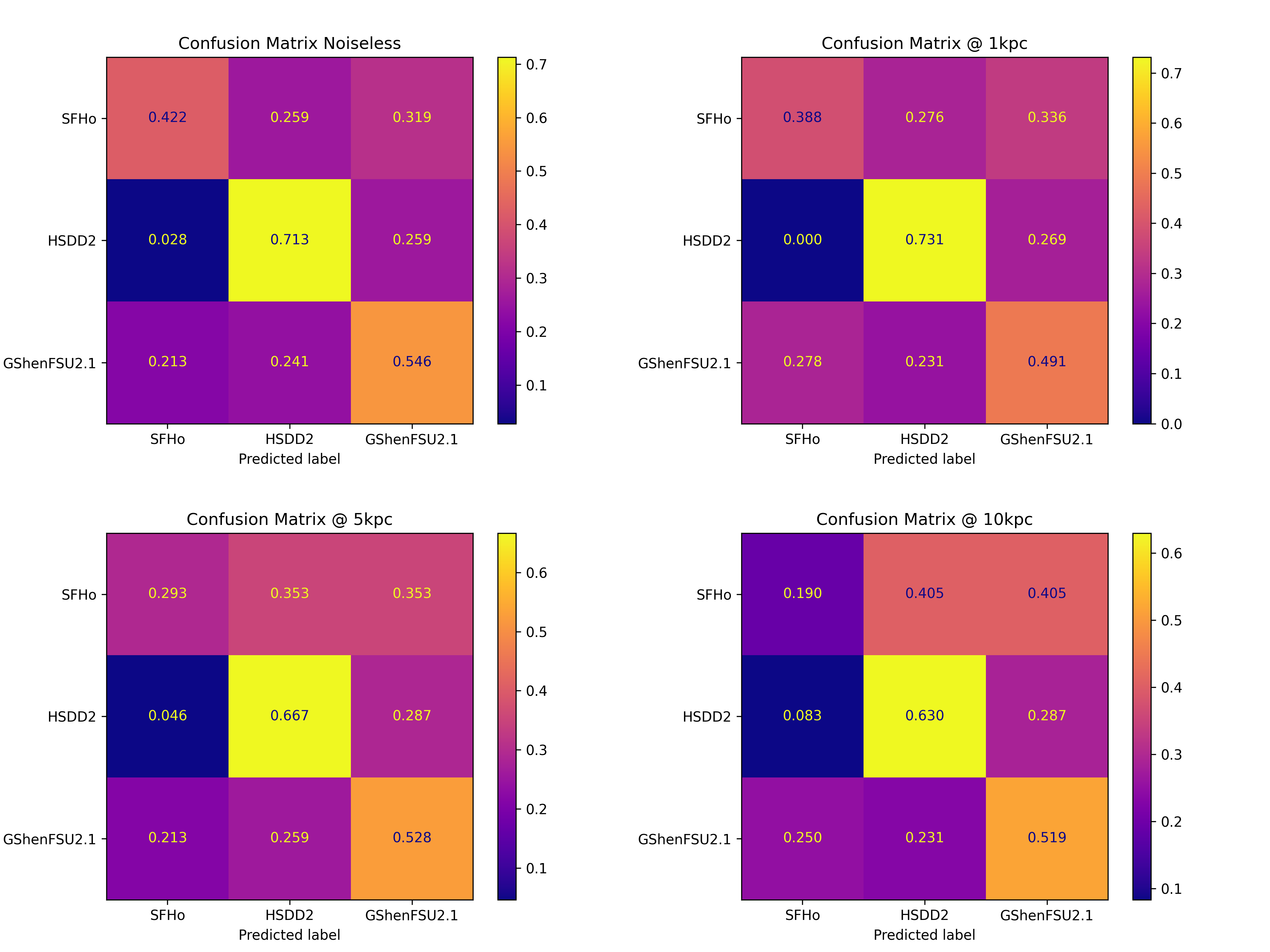}
    \caption{Confusion matrices for the EOS classification. The Noiseless case, 1kpc, 5kpc and 10 kpc.}
    \label{fig:confusion_matrices}
\end{figure*}

We proposed a simple EOS classificator based on residues $\varepsilon = |\alpha_{\text{est}} - \alpha_{\text{fit}}(\beta_{\text{est}})|$. We used 70\% of the Abylkairov signals for training, this is to calculate the curves coefficients and 30\% for testing. In figure \ref{fig:confusion_matrices} the confusion matrices are shown at three different distances and the noiseless case.

The individual accuracies were the following. For the Noiseless case: 0.762, 1kpc: 0.533, 5kpc: 0.491, 10kpc: 0.440. This low classification metric is related to the fact that the curves are grouped in two. The EOSs SFHo and GShenFSU2.1 are very similar to each other, let us called them the Group A. On the other hand, LS220 practically overlaps with HSDD2, let us call them Group B. This suggests similarity between the EOS in each group, and if we calculate the accuracies per group we have, for the Noiseless case: 0.881, 1kpc: 0.741, 5kpc: 0.684, 10kpc: 0.663. The apparent higher accuracy for 1kpc is a consequence of using all the waveforms from Abylkairov catalog instead just the 30\% for testing in the noiseless case.

\section{Conclusions}
\label{sec:ocho}

 This work extended an existing analytical model \cite{laura} for the GW core bounce component from RR CCSNe progenitors. We found that allowing a new timescale parameter '$s$' to vary leads to an improvement in the median fitting factor to waveforms from Richers Catalog, from 88.88\% to 90.83\%.  
 For the Richers database \cite{richers2017equation}, the rotational $\beta$ parameter was recovered in real interferometric noise with an average $ \sigma_{\beta} = 1.29\times 10^{-3}$ at 10 kpc, compared to previous results using matched filtering and maximum likelihood estimate (MLE) and real interferometric noise \cite{laura}, in which a value of $\sigma_\beta = 1.46\times 10^{-2}$ is reported. Those results, are consistent with the variances found in Gaussian colored noise by \cite{pastor2024bayesian}.
 Also, for the Abylkairov database, a median absolute relative error of 11.93\% with a 95th percentile of 38.41\% at 10 kpc, in real noise. Also using real noise, the case of prior sensitivity shows in the right panel of figure \ref{fig:O3KDE}, that the bias is higher for the three estimated parameters.
  Our analysis shows that the bias induced by different priors on the posterior PDFs to be less than 5 percent up to 10 Kpc for Gaussian colored noise and in real O3 noise we have a minimum bias of 0.6\% for triangular prior and an 11.9\% as a maximum bias in the $\beta$ parameter at 10 kpc. This analysis also shows from table \ref{tab:prior_bias_summary} and figure \ref{fig:O3KDE} that parameter $s$ is the most sensible to prior selection at 10 kpc.

 The process for injecting noise was to take a random time window around gps time (1256677376) for O3 observing run to draw different noise realizations. 
 As can be observed in figure \ref{fig:O3KDE}, for the noiseless case, the parameters are recovered with no bias. This study suggests that Bayesian PE on CCSNe, where no prior detections are available, should be paired with robustness studies to the choice of the priors.

We introduced a novel methodology for
estimating the nuclear EOS using a function of the estimated $\beta$ and $\alpha$. 
Explicitly, we found that  for $\beta <0.08$ different EOS lie along different curves in the $\alpha-\beta$ plane. As expected the classification performance degrades for increasing distances. We expect that considering a network of interferometers improves our estimation results for a distance $D$ by a factor of $\sqrt{N}$ where $N$ is the number of detectors. This study shows that Bayesian inference is an adequate tool for PE with physically motivated waveform models and real interferometric noise. This methodology can be considered for future GW searches and work as a foundation for extending this framework to future interferometers networks. In particular, the results presented here at 1 Kpc will be representative of 10 kpc for the Cosmic Explorer and Einstein Telescope.

The Code used for Parameter Estimation and results analysis is available at: \url{https://git.ligo.org/emmanuelalejandro.avila/bacco}.
%
\begin{acknowledgements}

E. A. and C.M. want to thank SNII-SECIHTI. M.Z. is supported by the National Science Foundation Gravitational Physics Experimental and Data Analysis Program through awards PHY-2110555 and PHY-2405227. Computational resources were provided by the LIGO Data Grid, supported by the U.S. National Science Foundation.
\end{acknowledgements}

\appendix

\section{Derivation of the relation between optimal and randomly oriented SNRs}
\label{sec:apendice}
The signal-to-noise ratio is defined as
\begin{equation}
\mathrm{SNR} \equiv \rho ^2 = (h|h),
\end{equation}
where the inner product is given by
\begin{equation}
(a|b) = 4 \text{Re} \left[ \int_0^\infty \frac{\tilde a(f)\tilde b^*(f)}{S_n(f)}\, df \right].
\end{equation}

The gravitational-wave strain can be written as
\begin{equation}
h(t) = F_+(\theta,\phi,\psi)\, \hbar_+(t;\lambda)
      + F_\times(\theta,\phi,\psi)\, \hbar_\times(t;\lambda),
\end{equation}
where $\lambda$ denotes intrinsic source parameters.

For a rotating triaxial ellipsoid \cite{creighton2012gravitational},
\begin{align}
\hbar_+(t) &= H_+(\iota)\, h_\times(t,\lambda), \\
\hbar_\times(t) &= H_\times(\iota)\, h_+(t,\lambda),
\end{align}
with
\begin{equation}
H_+(\iota) = \frac{1}{2}(1+\cos^2\iota), \qquad
H_\times(\iota) = \cos\iota,
\end{equation}
and
\begin{equation}
h_+(t) = \lambda \cos 2\omega t, \qquad
h_\times(t) = \lambda \sin 2\omega t,
\end{equation}
where $\lambda$ is a constant.

The SNR becomes
\begin{align}
(h|h) &= \left(
F_+ h_+ + F_\times h_\times \,\middle|\,
F_+ h_+ + F_\times h_\times
\right) \\
&= F_+^2 (h_+|h_+)
+ 2F_+F_\times (h_+|h_\times)
+ F_\times^2 (h_\times|h_\times).
\end{align}

Since $F_+$ and $F_\times$ are independent,
\begin{align}
\langle \rho^2 \rangle
&= \langle F_+^2\rangle \langle H_+^2\rangle (h_+|h_+)
+ 2\langle F_+F_\times\rangle \langle H_+H_\times\rangle (h_+|h_\times) \\
&\quad + \langle F_\times^2\rangle \langle H_\times^2\rangle (h_\times|h_\times).
\end{align}

The cross term vanishes,
\begin{equation}
(h_+|h_\times) = 0,
\end{equation}
since
\begin{equation}
\lambda^2 \int \cos 2\omega t \sin 2\omega t \, dt = 0
\end{equation}
by orthogonality.

The definition of the average is taken in accordance to the probability density function \cite{jaranowski2009analysis}:

\begin{equation}
  \label{eq:pdf}
  p(\iota,\theta,\phi,\psi) = \frac{1}{16\pi^2} \sin \iota \sin \theta
\end{equation}

Therefore,
\begin{equation}
\langle \rho^2 \rangle
= \langle F_+^2\rangle \langle H_+^2\rangle (h_x|h_x)
+ \langle F_\times^2\rangle \langle H_\times^2\rangle (h_x|h_x).
\end{equation}

The inclination averages are
\begin{align}
\langle H_+^2 \rangle
&= \int_0^\pi \frac{1}{4}(1+\cos^2\iota)^2 \sin\iota \, d\iota
= \frac{14}{15}, \\
\langle H_\times^2 \rangle
&= \int_0^\pi \cos^2\iota \sin\iota \, d\iota
= \frac{2}{3}.
\end{align}

The antenna pattern average is
\begin{widetext}
\begin{align}
\langle F_+^2 \rangle
&= \frac{1}{8\pi^2}
\int_0^\pi \int_0^{2\pi} \int_{0}^{\pi}
\frac{(1+\cos^2\theta)^2}{4}
\sin\theta \cos^2 2\phi \cos^2 2\psi + \cos^2 \theta \sin \theta \sin^2 2\phi \sin^2 2\psi
\, d\psi\, d\phi\, d\theta.
\end{align}
\end{widetext}

The mixed term vanishes since
\begin{equation}
\int_0^\pi \cos 2\psi \sin 2\psi \, d\psi = 0.
\end{equation}

Using
\begin{equation}
\int_{0}^{\pi} \sin^2 2\psi \, d\psi = \int_{0}^{\pi} \cos^2 2\psi \, d\psi = \frac{\pi}{2}, 
\end{equation}
one obtains
\begin{equation}
\langle F_+^2 \rangle
= \frac{1}{8\pi^2}
\left[
\frac{14}{15}\pi \cdot \frac{\pi}{2}
+ \frac{2}{3}\pi \cdot \frac{\pi}{2}
\right]
= \frac{1}{8}\left(\frac{24}{15}\right)
= \frac{3}{15}
= \frac{1}{10}.
\end{equation}

In the same way, we can calculate that:

\begin{equation}
  \label{eq:Fx}
  \langle F^2_{\times}\rangle = \frac{1}{10}.
\end{equation}

And,

\begin{equation}
  \label{eq:avg_snr}
  \langle \rho ^2 \rangle = \frac{14}{15}\cdot\frac{1}{10}\left(h_+|h_+\right) + \frac{2}{3}\cdot\frac{1}{10}\left(h_{\times} |h_{\times}\right).
\end{equation}

The inner products for the case of an ellipsoid involve taking the inner products of $\cos(2\omega t) \sin (2\omega t)$, since they differ only in phase:

$$
\left(h_+|h_+\right) = \left(h_\times|h_\times\right),
$$

$$
\langle \rho^2 \rangle = \frac{4}{25}\left(h_+|h_+\right).
$$

Optimal SNR means that the line of sight is aligned with the rotation axis of the radiating source, this is $\iota = 0$ and $H_{+} = H_\times = 1$. The antenna patterns are $F_+ = 1$ and $F_\times = 0$ which means that the detector observes completely just one polarization.

\begin{equation}
  \label{eq:opt_snr}
  \langle \rho^2_{\text{opt}} \rangle = \langle F^2_{+}\rangle \langle H^2_+\rangle\left(h_+|h_+\right) = \left(h_+|h_+\right).
\end{equation}

The fraction, then is:

\begin{equation}
  \label{eq:fraction}
  \sqrt{\frac{\langle \rho^2\rangle}{\langle \rho^2_{\text{opt}} \rangle}} = \frac{2}{5}. 
\end{equation}

\newpage
\appendix

\newpage

\bibliographystyle{unsrt}
\bibliography{References}
\end{document}